\documentclass[aip,apl,reprint]{revtex4-1}
\usepackage{amsmath,graphicx,float}
\usepackage{siunitx,mathtools,amssymb}
\usepackage{subfigure} 
\usepackage[usenames]{color}
\usepackage{dcolumn}%
\usepackage{bm}%
\usepackage{textcomp} 
\usepackage[resetlabels]{multibib}

\begin{document} 
 
\title {\color{blue} High Kinetic Inductance Microwave Resonators Made by He-Beam Assisted Deposition of Tungsten Nanowires}
\author{J. Basset} 
\affiliation{Laboratoire de Physique des Solides, CNRS, Universit\'e  Paris-Sud,
	Universit\'e Paris Saclay, 91405 Orsay Cedex, France.}  

\author{D. Watfa} 
\affiliation{Laboratoire de Physique des Solides, CNRS, Universit\'e  Paris-Sud,
	Universit\'e Paris Saclay, 91405 Orsay Cedex, France.} 

\author{G. Aiello} 
\affiliation{Laboratoire de Physique des Solides, CNRS, Universit\'e  Paris-Sud,
	Universit\'e Paris Saclay, 91405 Orsay Cedex, France.}

\author{M. F\'echant} 
\affiliation{Laboratoire de Physique des Solides, CNRS, Universit\'e  Paris-Sud,
	Universit\'e Paris Saclay, 91405 Orsay Cedex, France.}

\author{A. Morvan} 
\affiliation{Laboratoire de Physique des Solides, CNRS, Universit\'e  Paris-Sud,
	Universit\'e Paris Saclay, 91405 Orsay Cedex, France.}

\author{J. Esteve} 
\affiliation{Laboratoire de Physique des Solides, CNRS, Universit\'e  Paris-Sud,
	Universit\'e Paris Saclay, 91405 Orsay Cedex, France.}

\author{J. Gabelli} 
\affiliation{Laboratoire de Physique des Solides, CNRS, Universit\'e  Paris-Sud,
	Universit\'e Paris Saclay, 91405 Orsay Cedex, France.}

\author{M. Aprili} 
\affiliation{Laboratoire de Physique des Solides, CNRS, Universit\'e  Paris-Sud,
	Universit\'e Paris Saclay, 91405 Orsay Cedex, France.}

\author{R. Weil} 
\affiliation{Laboratoire de Physique des Solides, CNRS, Universit\'e  Paris-Sud,
	Universit\'e Paris Saclay, 91405 Orsay Cedex, France.}

\author{A. Kasumov} 
\affiliation{Laboratoire de Physique des Solides, CNRS, Universit\'e  Paris-Sud,
	Universit\'e Paris Saclay, 91405 Orsay Cedex, France.}

\author{H. Bouchiat} 
\affiliation{Laboratoire de Physique des Solides, CNRS, Universit\'e  Paris-Sud,
	Universit\'e Paris Saclay, 91405 Orsay Cedex, France.}  

\author{R. Deblock} 
\affiliation{Laboratoire de Physique des Solides, CNRS, Universit\'e  Paris-Sud,
	Universit\'e Paris Saclay, 91405 Orsay Cedex, France.} 

\begin{abstract}

We evaluate the performance of hybrid microwave resonators made by combining sputtered Nb thin films with Tungsten nanowires grown with a He-beam induced deposition technique. 
Depending on growth conditions the nanowires have a typical width $w\in[35-75]$~nm and thickness  $t\in[5-40]$~nm. We observe a high normal state resistance  $R_{\ensuremath{\Box}}\in [65-150]$ $\Omega/\ensuremath{\Box}$ which together with a critical temperature $T_c\in[4-6]~K$ ensure a high kinetic inductance making the resonator strongly nonlinear.
Both lumped and coplanar waveguide resonators were fabricated and measured at low temperature exhibiting internal quality factors up to $3990$ at $4.5$~GHz in the few photon regime. Analyzing the wire length, temperature and microwave power dependence we extracted a kinetic inductance for the W nanowire of $L_K\approx15$ pH/$\ensuremath{\Box}$, which is 250 times higher than the geometrical inductance, and a Kerr non-linearity as high as $K_{W,He}/2\pi=200 \pm 120$~Hz/photon at $4.5$~GHz. The nanowires made with the helium focused ion beam are thus versatile objects to engineer compact, high impedance, superconducting environments with a mask and resist free direct write process.

\end{abstract}

\maketitle  
High kinetic inductance superconducting materials are having a growing impact in the circuit quantum electrodynamics community. Operated in the microwave frequency domain they allow to engineer high impedance circuits providing an efficient way to increase the lifetime of superconducting quantum bits \cite{Manucharyan2009,Pop2014,Lin2017,Earnest2018}, couple electron charge and spins to microwave photons\cite{Stockklauser17,Landig2018}, study the coherent quantum phase slip\cite{Astafiev2012} or generate a high impedance environment in dynamical Coulomb blockade experiments \cite{Altimiras13,Rolland2018}. These materials may consist of arrays of Josephson junction\cite{Pop2014,Altimiras13}, disordered thin films of metallic compounds such as NbTi \cite{McCambridge95}, NbN \cite{Gao07,Niepce18},NbSi \cite{Calvo14},TiN\cite{Leduc10,Shearrow18},NbTiN \cite{Barends08,Samkharadze16}, granular aluminium \cite{Maleeva18} or superconducting semiconductors \cite{Shim2015}. Their use in superconducting circuits usually requires etching and/or final sharpening with electron-beam lithography. By contrast beam-assisted deposition of superconducting materials allows to design and deposit extremely narrow superconducting nano-objects with a versatile direct write process with no templates such as masks or resists. This could  be a crucial point to incorporate fragile and delicate samples into a resonant structure.  Until now, tungsten superconducting nanowires have been deposited using Ga$^+$ ions in a Focused Ion Beam (FIB)\cite{Sadki04,Kasumov05,Sadki05} or electrons in an electron scanning microscope (SEM) \cite{Sengupta15}. This tungsten has already been used to connect nanoscale samples  such as fullerenes \cite{Kasumov05}, graphene\cite{Shailos07}, mesoscopic metallic samples\cite{Chiodi12}, Bi nanowires\cite{Li14,Murani17} or study superconductivity in low dimension\cite{Guillamon09,Li11}. 
The focused helium ion beam from gas field-ion sources \cite{Ward06} allows to fabricate extremely narrow nanowires with potentially better superconducting properties than the ones realized with electron beam, and less damage and contamination than with a Ga-FIB. Connecting nanoscale sample with this technique is possible. However the high frequency properties, and especially the use of W nanowire as compact non-linear high impedance superconducting elements for quantum electronics,
has not yet been explored.

We show here that it is possible to grow superconducting W nanowires with the He-ion beam induced deposition (He-IBID) technique  that present attractive properties for high frequency superconducting circuits. The letter is organized as follows. After detailing the sample fabrication we report resonators datasets as a function of temperature and applied power from which we infer the kinetic inductance fraction and the non-linear Kerr parameters. We conclude with potential applications.

\begin{figure}[htbp]
	\begin{center}
		\includegraphics[width=8.5cm]{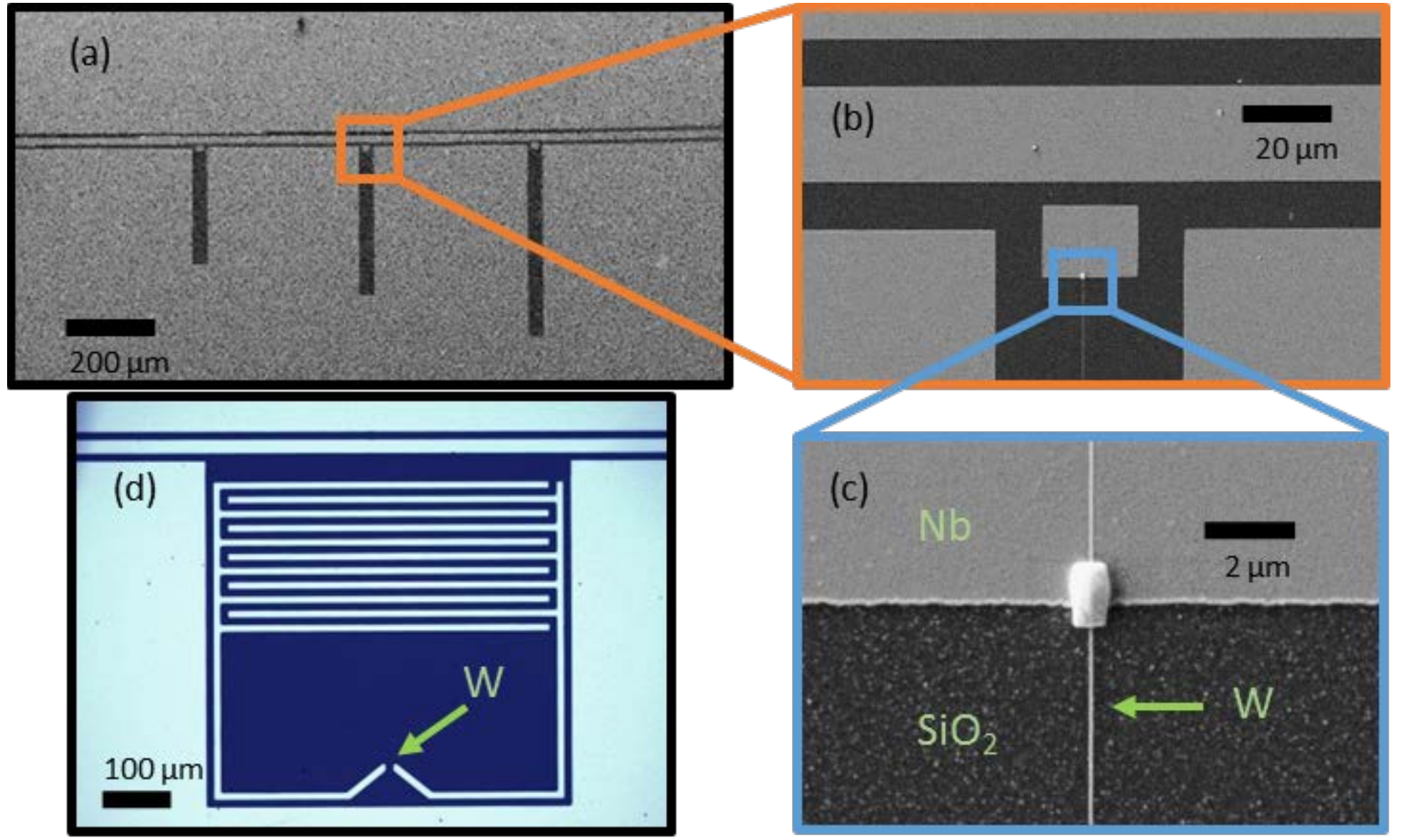}
	\end{center}
	\caption{(a) SEM image of a CPW sample. The transmission line runs horizontally with three $\lambda/4$ resonators hanging vertically. (b) SEM image of the coupling area between CPW resonator and transmission line. (c) SEM image of a W nanowire connected to Nb with a thick W patch at the junction. (d) Optical microscope picture of the lumped resonator. The arrow points the nanowire placed horizontally.}
	\label{Figure0}
\end{figure}

Two types of design were compared: a coplanar waveguide (CPW) resonator (Fig.~\ref{Figure0}a and b) and a lumped element resonator (Fig.~\ref{Figure0}d). Separate control samples with Titanium/Gold contacts were used to measure the nanowires dc properties. To fabricate the hybrid resonators a $110$~nm thick layer of Nb is sputtered on a high resistivity silicon substrate with a $500$~nm thick thermal oxide. We then use optical lithography to define a positive mask on top of Nb for reactive ion etching with SF$_6$.  The W nanowire is then deposited to form a resonator: a long grounded wire capacitively coupled to a transmission line for the $\lambda/4$ CPW resonator and in parallel with an interdigitated capacitor in the lumped geometry (Fig.~\ref{Figure0}c and d).

The W nanowires are fabricated using helium beam assisted deposition in an ORION Nanofab microscope from Zeiss company. It consists of an ultra-high brightness gas field ionization source (GFIS). The source is formed by three atoms at the apex of a metallic tip cryogenically cooled with Nitrogen and submitted to a high voltage ($30$~kV) in the presence of helium gas. An ionized gas is generated at the end of the tip and emission originates preferentially from the apex. Only He ions from one atom of the trimer are selected for imaging and nanofabrication. The extracted ions are accelerated by the column of the microscope to raster scan as in a SEM. Imaging is done via secondary electrons collected into an Everhardt-Thornley detector. During nanodeposition we use a He-beam current ranging from $10$ to $30$~pA. The W precursor (tungstenhexacarbonyl W(CO)$_6$) is injected by a gas injection system OmniGIS II from Oxford Instruments. The target pressure in the chamber is $4\cdot10^{-6}$~Torr. The patterning geometry and parameters are controlled by the nanofabrication system NPVE from Fibics. 
Table \ref{tableDC} shows the dimensions and growth conditions of the nanowires obtained using the above-described procedure. Heights were measured by atomic force microscopy and widths by SEM.
For the lumped element resonator a W nanowire of $9.8\mu$m was realized whereas for the CPW $\lambda/4$ resonator a length of $390\mu$m was used. Fabricating such a long nanowire is possible thanks to the very high stability of the He-FIB and He-IBID process compared to previous experiments with Ga-FIB\cite{Kasumov05}.
The quality of the W nanowires growth was controlled by depositing tungsten nanowires on samples dedicated to electrical dc measurements. These control samples have been measured at low temperatures (see appendix \ref{AppendixA}) and the dc properties of three samples are summarized in table \ref{tableDC}. They exhibit a superconducting transition with critical temperature $T_c\in[5-6.5]$~K and a critical field $H_{c2}$ higher than 5T, similar to properties of 3D grown W nanowires\cite{Cordoba18}.

\begin{table*}
	\renewcommand\arraystretch{1.2}
	\begin{center}
		\begin{tabular}{|c||c|c|c|c|c|c|c|c|c|c|}
			\hline
			Sample & Current & Dose & Length & Thickness & Width & Resistance & $R_{\ensuremath{\Box}}$ & $\rho$ & $\xi(2K)$ & $\lambda(0K)$ \\
			& (pA) & & ($\mu$m) & (nm) & (nm) & (k$\Omega$) &($\Omega$) & ($\mu \Omega$.cm) & (nm) & (nm) \\
			\hline
			NW1 &	20 & 0.178nC/$\mu$m & 5.9 & 40 & 50 & 7.75 & 65.7 & 266 & 6.7 & 674\\
			NW2 &	20 & 0.06nC/$\mu$m & 5.9 & 5.5 & 35 & 25.15 & 149.2 & 80.6 & 7.6 & 400 \\
			NW3 & 10 & 3nC/$\mu$m$^{2}$ & 5.9 & 20 & 70 & 9.1 & 108.0 & 216.0 & 7 & 449\\
			Resonator "Lumped" & 20 & 0.16nC/$\mu$m & 9.8 & 30 & 35 & & & & &\\
			Resonator "$\lambda/4$" & 27 & 3nC/$\mu$m$^{2}$ & 390 & 12 & 75 & & & & &\\
			Resonator "$Lumped2$ " & 27 & 3nC/$\mu$m$^{2}$ & 30 & 25 & 80 & & & & &\\
			\hline
		\end{tabular}		
	\end{center}
	\caption{Growth parameters, dimensions and transport properties of the fabricated W nanowires. With the indicated parameters the time needed to write the W part of the resonator "Lumped" is 88 seconds , whereas for the "$\lambda/4$" type it is 40 minutes.}
\label{tableDC}
\end{table*}

The microwave characterization of the resonators allows extracting the kinetic inductance fraction of the W nanowire, its kinetic inductance per unit length and finally the non-linear Kerr parameters. 
The kinetic inductance can be estimated, at very low temperature, from $R_{\ensuremath{\Box}}$ and $T_C$ \cite{Annunziata2010}:
\begin{equation}
L_{K,\ensuremath{\Box}}\approx\frac{R_{\ensuremath{\Box}} h}{2\pi ^2 \Delta_0}.
\end{equation}
Taking $T_c\approx5.0$~K, $\Delta_0=1.76 k_B T_c$ and the square resistances from table \ref{tableDC} we evaluate a kinetic inductance $L_{K,\ensuremath{\Box}}\in [7-25]$~pH$/\ensuremath{\Box}$. Using this estimate we designed microwave resonators resonating in the $2-6$~GHz microwave range capacitively coupled to a transmission line (Fig.~\ref{Figure0}). Each sample is placed into a copper box equipped with a printed circuit board and thermally anchored to the $10$~mK stage of a dry dilution unit.  Microwaves are sent via attenuated and thermally anchored microwave lines. The transmitted wave is amplified and the complex transmission spectra $S_{21}$ through the line is measured with a vectorial network analyzer.

At low temperature the whole structure is superconducting. One probes the properties of the microwave resonator in a hanger type of geometry where the interference of the incident microwave signal and the one reflected from the resonator lead to a dip in $S_{21}$. This dip is accounted theoretically by:
\begin{equation}
S_{21}=1-\frac{Q_t}{Q_c} \frac{1-2 j Q_c u}{1+2jQ_tx}
\label{Formula0}
\end{equation}
with $x=(\omega - \omega_0)/\omega_0$ the fractional detuning of the readout angular frequency $\omega$ relative to the resonance frequency $\omega_0$. $u$ is a dimensionless parameter taking into account the asymmetry in the transmission line and is essential to extract reliable quality factors in hanger-coupled resonators\cite{Khalil2012}. It reduces to $0$ for a symmetric transmission line.
The coupling quality factor $Q_c$ quantifies the coupling between the transmission line and the resonator which has an intrinsic quality factor $Q_i$. These two terms are related to the total (or loaded) quality factor $Q_t$ via $Q_t^{-1}=Q_i^{-1}+Q_c^{-1}$. $Q_i$ gives information on the quality of the resonator independently of the coupling to the measurement line and is a figure of merit of the material quality. 
    
\begin{figure}[htbp]
	\begin{center}
		\includegraphics[width=8.5cm]{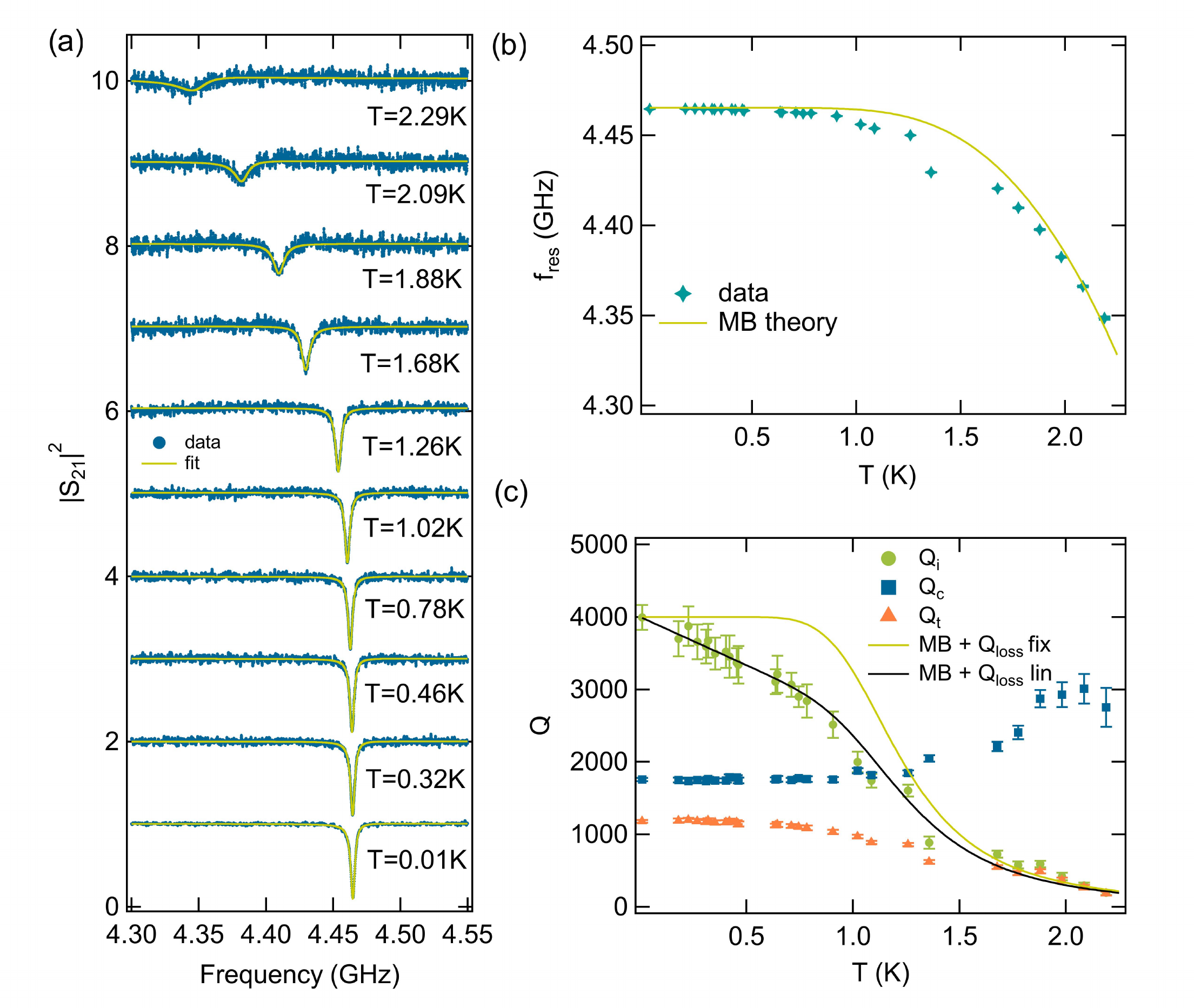}
	\end{center}
	\caption{(a) Temperature dependence of the normalized transmission spectra of the lumped resonator. (b) Temperature dependence of the resonance frequency. (c) Extracted quality factors \textit{vs} temperature.}
	\label{Figure2}
\end{figure}

The transmission spectra of the lumped resonator at $10$~mK and low power is shown in Fig.~\ref{Figure2}a with a fit using Eq.~\ref{Formula0}. We found $\omega_0/2\pi=4.4642$GHz and $Q_i=3990$ at low power. For comparison a resonator entirely made of Nb with a similar design exhibits $Q_i=11000$ at $T=1.6$~K. Using a finite element simulation using Sonnet$^{\mbox{\scriptsize{\textregistered}}}$ software we could extract the parallel capacitance $C\approx240$~fF, the geometrical inductance of the design (corresponding to the wide Nb inductive part) $L_{geo,D}\approx0.935$~nH and the kinetic inductance of the wire $L_{K}=4.3$~nH. The geometrical inductance of the nanowire alone is expected to be $L_{geo,W}$$\approx17$~pH leading to the kinetic inductance fraction $\alpha=L_K/(L_{geo,W}+L_K)=0.996$ and kinetic over geometrical inductance ratio $\beta=L_K/L_{geo,W}=253$ so that the geometrical inductance can be neglected. 
We obtain a kinetic inductance per unit length $\cal{L}$$_{K}$$\approx439\mu$H/m and a kinetic inductance per square $L_{K,\ensuremath{\Box}}=15.4$pH$/\ensuremath{\Box}$ which falls in the expectation window calculated earlier. These values are sizeable even compared to state of the art material like NbN ($L_{K,NbN,\ensuremath{\Box}}\in[4.4,82]$~pH$/\ensuremath{\Box}$\cite{Annunziata2010,Niepce18})  or NbSi thin films ($L_{K,NbSi,\ensuremath{\Box}}=20$~pH$/\ensuremath{\Box}$\cite{Calvo14}). The kinetic inductance of the W nanowire can be increased by reducing the thickness and/or width of the wire and less reliably by changing the growth condition. 

 The coplanar waveguide (CPW) $\lambda/4$ resonator gave a smaller quality factor $Qi\approx 710$ (appendix \ref{AppendixB}). In this geometry the wire length was $390$~$\mu$m long with a resonance frequency $\omega_0/2\pi=4.05$~GHz. From the design we extracted the lineic capacitance to ground to $\cal{C}$$\approx48$pF/m and the lineic geometrical inductance of the wire $\cal{L}$$_{geo}$$\approx1.7\mu$H/m. The lineic inductance was then $\cal{L}$$=$$\cal{L}$$_{geo}+$$\cal{L}$$_{K}$$\approx512\mu$H/m.  
From these numbers we extracted a phase velocity $c=1/\sqrt[]{\cal{L}\cal{C}}=6.4\times10^6$~m/s and a characteristic impedance $Z_C=\sqrt[]{\cal{L}/\cal{C}}=3.3$ k$\Omega$. Such material is therefore highly suitable for dynamical Coulomb blockade experiments where the characteristic impedance $Z_C$ must be comparable to the resistance quantum $R_Q=h/4e^2\approx 6.5 k\Omega$. More specifically, the coupling of $e.g.$ a tunnel junction to a high impedance microwave resonator is characterized by the coupling parameter $\lambda=\sqrt{\pi Z_C/R_Q}$. With $\lambda\approx1.26$ in our experiment we would be at the onset of the strong coupling regime $\lambda\geqslant 1$ where e.g. dc-driven single microwave photon generation could be achieved\cite{Souquet2016,Esteve2018}. 

Figure \ref{Figure2}a shows transmission spectra of the lumped resonator as a function of temperature with fits using equation \ref{Formula0}. Decreasing the temperature leads to an increase of the resonance frequency together with a sharpening of the resonance (Fig.~\ref{Figure2}b and c). Below $T=0.9$K the resonance frequency is nearly constant and decreases strongly as one raises the temperature. $Q_i$ evolves similarly with a maximum value reaching $Qi=3990$ at $10$mK.

We use the Mattis-Bardeen (MB) theory \cite{Mattis1958,Barends08} (appendix \ref{AppendixB}) as a minimal model to relate the temperature dependence of the resonance frequency and quality factor to the complex conductivity of the material $\sigma=\sigma_1-i\sigma_2$. We make the crude assumption that only $\sigma$ of the nanowire is relevant. This is justified by the fact that the ac current density is much higher in the nanowire compared to the Nb circuit which has much larger transverse dimensions and is expected to exhibit smaller losses. Let us emphasize here that the MB theory is not directly applicable for very disordered superconductors due to rounded BCS density of states and inhomogeneities \cite{Ghosal1998,Sacepe2008,Driessen2012,Coumou2013}, but constitutes a first attempt to model our samples. 

The results are shown along with the experimental data in Fig.~\ref{Figure2}b and c. Assuming a zero-temperature superconducting gap $\Delta_0\approx750$~$\mu$eV this theory allows to reasonably tackle the temperature dependence of the resonance frequency whereas discrepancies are found regarding the quality factors for which the MB theory predicts a diverging $Q_i$ as the temperature is lowered. 
To reproduce our data we introduce extra losses in the internal $Q_i$ factor as:
$Q_i^{-1}= Q_{MB}^{-1}+Q_{loss}^{-1}$.
A temperature independent $Q_{loss}=3990$ leads to the yellow trace in Fig.~\ref{Figure2}c. Whereas the saturation value is fine, the global agreement in the intermediate temperature regime is unsatisfactory. To reproduce best our data a phenomenological linear decay of $Q_{loss}$ with respect to temperature defined as $Q_{loss}=3990(1-T/3)$ was introduced. The corresponding curve is shown in black in figure \ref{Figure2}c demonstrating a reasonable  agreement. This decay sheds light on the physics of losses into the resonator which increases as the temperature is raised. These extra losses need to be further understood and may be related to poisoning\cite{Oates1991,Zemlicka2015,Maleeva18}, TLS \cite{Samkharadze16,LeSueur2018} and/or mobile vortices\cite{Tinkham1996}.

Because of the high critical fields of both materials used for this experiment, Nb and W, the resonator should be rather immune to applied magnetic field. We measured the properties of a lumped element resonator as a function of an applied in plane magnetic field up to $130$~mT and observed only small changes (see appendix \ref{AppendixB}, $<10\%$ in quality factor and $<0.05\%$ in resonance frequency.). This makes this hybrid resonator interesting to study mesoscopic devices where the spin degree of freedom needs to be addressed\cite{Viennot2015,Samkharadze16,Landig2018}.

Increasing the microwave power results in the onset of nonlinear effects due to the nonlinear kinetic inductance of the nanowire. 
This leads to the behaviour of a Duffing oscillator (appendix \ref{AppendixB}). To quantitatively tackle this non linear effect one needs to take into account the shift $\delta\omega$ of the resonance frequency due to the non linear kinetic inductance. The shifted resonance reads\cite{Swenson2013}
$\omega_r=\omega_0+\delta\omega=\omega_0+K n_{ph}$
where we have introduced the Kerr parameter $K$ relating the frequency shift to the number of photons stored in the resonator:

\begin{equation}
n_{ph}=\frac{2 Q_t^2 P}{Q_c \hbar \omega_0^2} \frac{1+4\frac{Q_c^2 Q_t}{Q_c-Q_t} u x}{1+4 Q_t^2 x^2}.
\end{equation}
$P$ is the power at the input of the resonator and
\begin{equation}
x= x_0-\frac{K n_{ph}}{\omega _0}= x_0 - K \frac{2 Q_t^2 P}{Q_c \hbar \omega_0^3} \frac{1+4\frac{Q_c^2 Q_t}{Q_c-Q_t} u x}{1+4 Q_t^2 x^2}.
\label{newdefxbis}
\end{equation}
Introducing $y=Q_t x$, $y_0=Q_t x_0$, $v=Q_t u$, the non linear parameter 
\begin{equation}
a_{NL}= -\frac{2 K Q_t^3}{Q_c \hbar \omega_0^3} P
\label{aNL}
\end{equation}
and defining $c=\frac{Q_c^2 Q_t}{Q_c-Q_t}$ we can rewrite the expression \ref{newdefxbis} as
$y=y_0 + \frac{a[1+4cvy]}{1+4y^2}$.
This expression takes into account the asymmetry of the transmission line. It consists in a third degree polynomial with three solutions post-selected to match the measurement sweep direction. By solving this equation we calculate the power-dependent frequency detuning $x$ and fit the experimental data using Eq.~\ref{Formula0}. Doing so allows one to plot as a function of the applied microwave power $P$, the non linear term $a_{NL}$, $Q_i$, $Q_c$ and $Q_t$. As expressed in \ref{aNL}, $a_{NL}$ should be a linear function of the power with a slope $s$ related to the non-linear Kerr parameter $K$ such that $K=-\frac{Q_c \hbar \omega_0^3}{2 Q_t^3} s$.
\begin{figure}[htbp]
	\begin{center}
		\includegraphics[width=8.5cm]{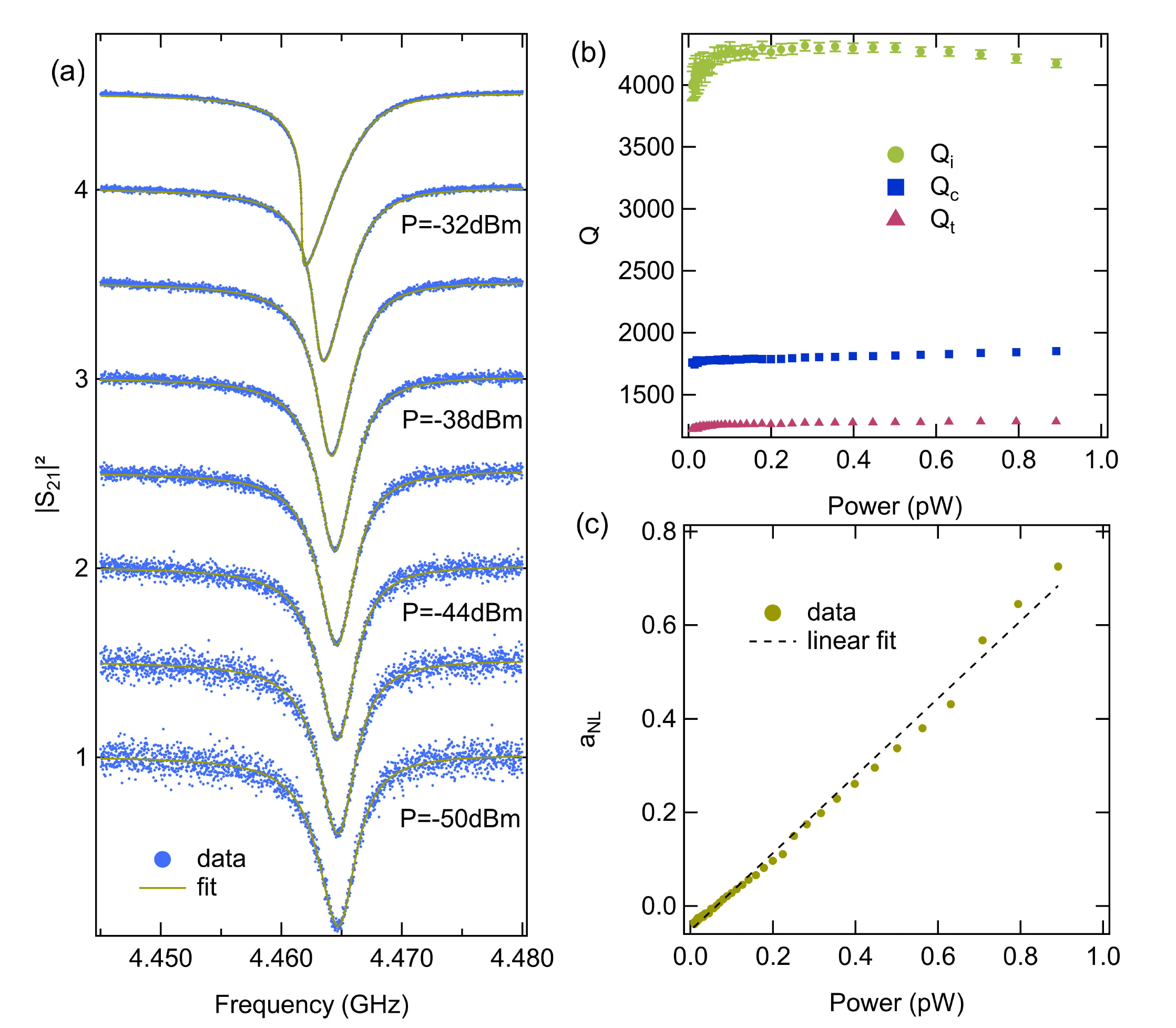}
	\end{center}
	\caption{(a) Power dependence of the normalized transmission spectra of the lumped resonator at $10$~mK.(b) Extracted quality factors \textit{vs} microwave power. (c) Power dependence of the $a_{NL}$ term (see text). }
	\label{Figure3}
\end{figure}

We have measured the power dependence of the transmission spectra of the resonator. The data together with the results of the fit are shown in Fig.~\ref{Figure3}. Figure \ref{Figure3}a shows that as one increases the microwave power, the peak shifts towards smaller frequency, slants and eventually becomes bistable at high power. From the fit we observe that the quality factors (Fig.~\ref{Figure3}b) slightly increase from $Q_i=3990$ to $4290$ as the power is elevated up to $0.1$~pW and remains stable at higher powers. This may indicate the saturation of two-level systems but should lead to much higher increase of quality factor\cite{Sage2011}. The $a_{NL}$ parameter (Fig.~\ref{Figure3}c) is linear with respect to microwave power P. 
From the slope $s$ we extract a Kerr parameter $K/2\pi=200 \pm 120$~Hz/photon at $4.465$~GHz, the uncertainty on $K$ being due to the uncertainty on the microwave power P reaching the sample.

To conclude we evaluated the performance of hybrid microwave resonators made by combining sputtered Nb thin films with Tungsten nanowires grown with a He-beam induced deposition technique. 
Both lumped and coplanar waveguide resonators were fabricated and measured at low temperature. They exhibit internal quality factors at high power up to $Qi=4290$ at $4.46$~GHz for $T=10$~mK. We extracted a large kinetic inductance for the W nanowire of $L_K=15.4$ pH/$\ensuremath{\Box}$, a kinetic inductance fraction $\alpha_{W,He}=0.996$ and a Kerr non-linearity of the resonators $K_{W,He}/2\pi=200 \pm 120$~Hz/photon at $4$~GHz. We also verified (appendix \ref{AppendixB}) that the resonators are immune to in-plane magnetic field up to $130$~mT the highest field achievable in our experimental setup. As such, this hybrid resonator could be interesting to study mesoscopic devices where the spin degree of freedom needs to be addressed\cite{Viennot2015,Landig2018}.
Finally this study allowed us to conclude that nanowires made with the He-FIB are versatile tools to engineer compact, high impedance, superconducting environments with a direct-write and resist-free process. It could prove useful to detect optical or plasmonic light or study the dynamical Coulomb blockade.

We acknowledge valuable discussions with F. Chiodi, H. Lesueur, P. Joyez and S. Gu\'eron. This work has been funded by the CNRS, CEA, University Paris-Sud, Paris Ile-de-France Region in the framework of DIM SIRTEQ and DIM NANO’K, Labex PALM (ANR-10-LABX-0039-PALM), Labex NANOSACLAY, Lidex NANODESIGN and the French ANR JETS (ANR-16-CE30-0029-01) and ANR INTELPLAN (ANR-15-CE24-0020).

\begin{appendix}
	
\section{dc-characterization of the W nanowires}
\label{AppendixA}
\begin{figure}[]
	\begin{center}
		\includegraphics[width=8.5cm]{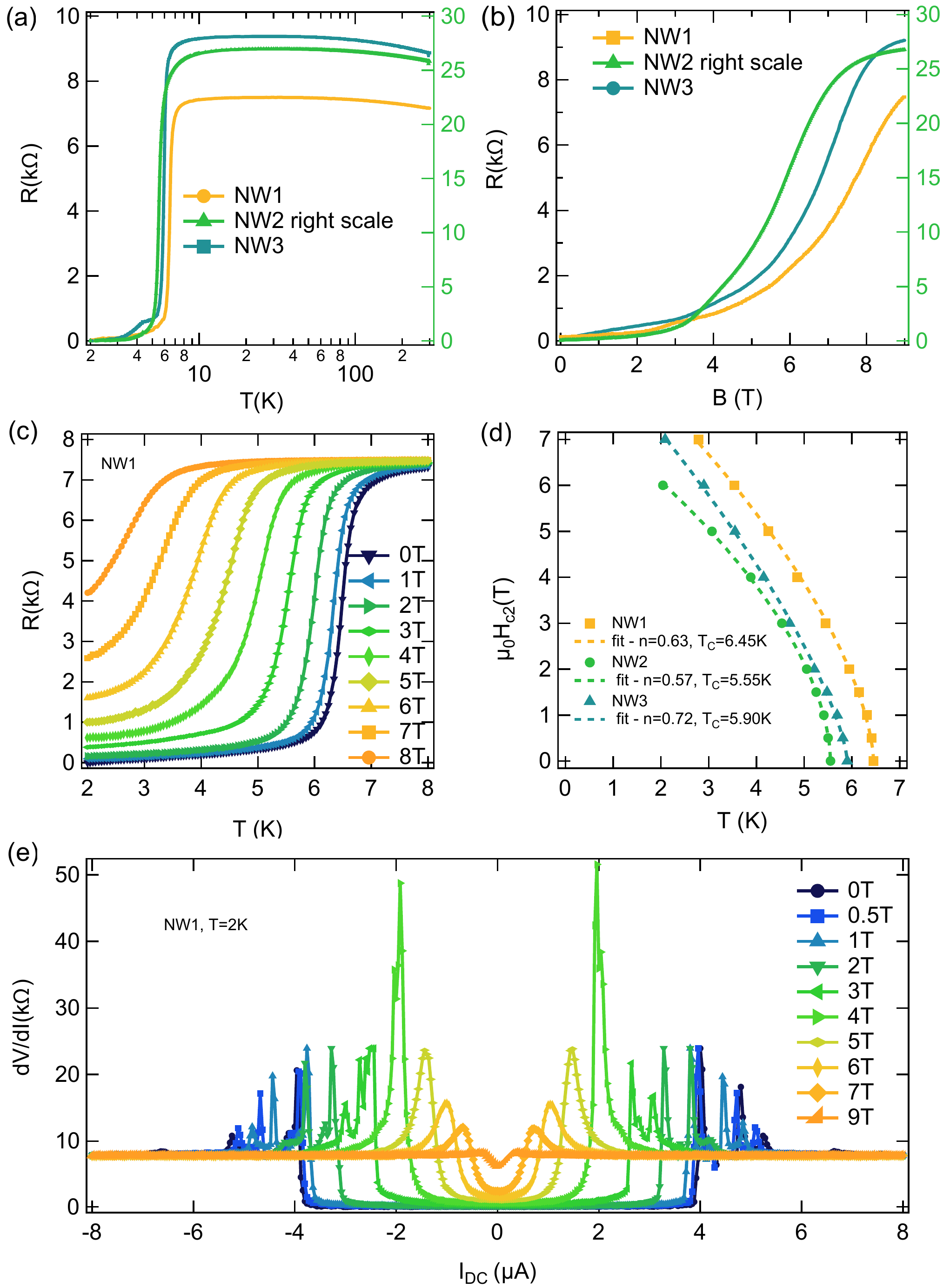}
	\end{center}
	\caption{(a) Resistance of W nanowires referenced in table 1 of the article \textit{vs} temperature $T$ and (b) perpendicular magnetic field $B$. (c) Resistance of sample NW1 \textit{vs} $T$ for different $B$. (d) Critical field $H_{C2}$ \textit{vs} $T$ of the wires with the corresponding fit (see text). Fitting parameters are indicated in the legend. (e) Differential resistance \textit{vs} current of sample NW1 for different magnetic fields.}
	\label{FigureDC}
\end{figure}

The quality of the W nanowires growth was controlled by depositing tungsten nanowires on samples dedicated to electrical dc measurements. These samples have been measured at low temperatures in a commercial variable temperature insert. We show in figure \ref{FigureDC} a and b the typical evolution of the resistance\cite{footnote} as a function of the temperature and magnetic field for three samples differing by their growth conditions (see table 1 in the article) and labeled NW1,NW2 and NW3. 
As one lowers the temperature the resistance increases slightly ($\approx5$\%) to finally show a superconducting transition around $T_c\in[5-6.5]$K. At $2$~K the nanowires are superconducting with a resistance that develops a perpendicular magnetic field behaviour consistent with a type $2$ superconductor. Defining Hc$_2$ as the magnetic field for which the nanowire recovers half of its normal state resistance we see that all wires exhibit an $H_{c2}$ larger than $5$~T. 

We further show in figure \ref{FigureDC}c the temperature dependence of the resistance of the nanowire NW1 for different perpendicular magnetic fields. The temperature dependence of $H_{c2}$ is plotted in figure \ref{FigureDC}d for the three wires. Such data can be fitted by the power law $H_{c2}(T)\propto (1-T/T_c)^n$ with $n$ ranging from $0.57$ to $0.72$ depending on the sample (see figure \ref{FigureDC}d). For a purely $2D$ superconductor $n=1$ is expected for perpendicular magnetic fields and $n=0.5$ for parallel magnetic fields. Finding an intermediate value of $n$ points towards reduced dimensionality of superconductivity in the W nanowires\cite{Qin2018,Cordoba18,Makise2016}.
From the value of $H_{C2}$ and $T_C$ we extract the superconducting coherence lengths $\xi$ and London penetration lengths $\lambda$,  reported in table 1 in the article, assuming\cite{Tinkham1996} $\mu_0 H_{C2}(T)=\Phi_0/2\pi \xi^2(T)$  and $\lambda (0)=1.05 10^{-3} \sqrt{\rho/T_c}$ with $\rho$ the resistivity\cite{Kes1983}. 

\begin{figure}[htbp]
	\begin{center}
		\includegraphics[width=8.5cm]{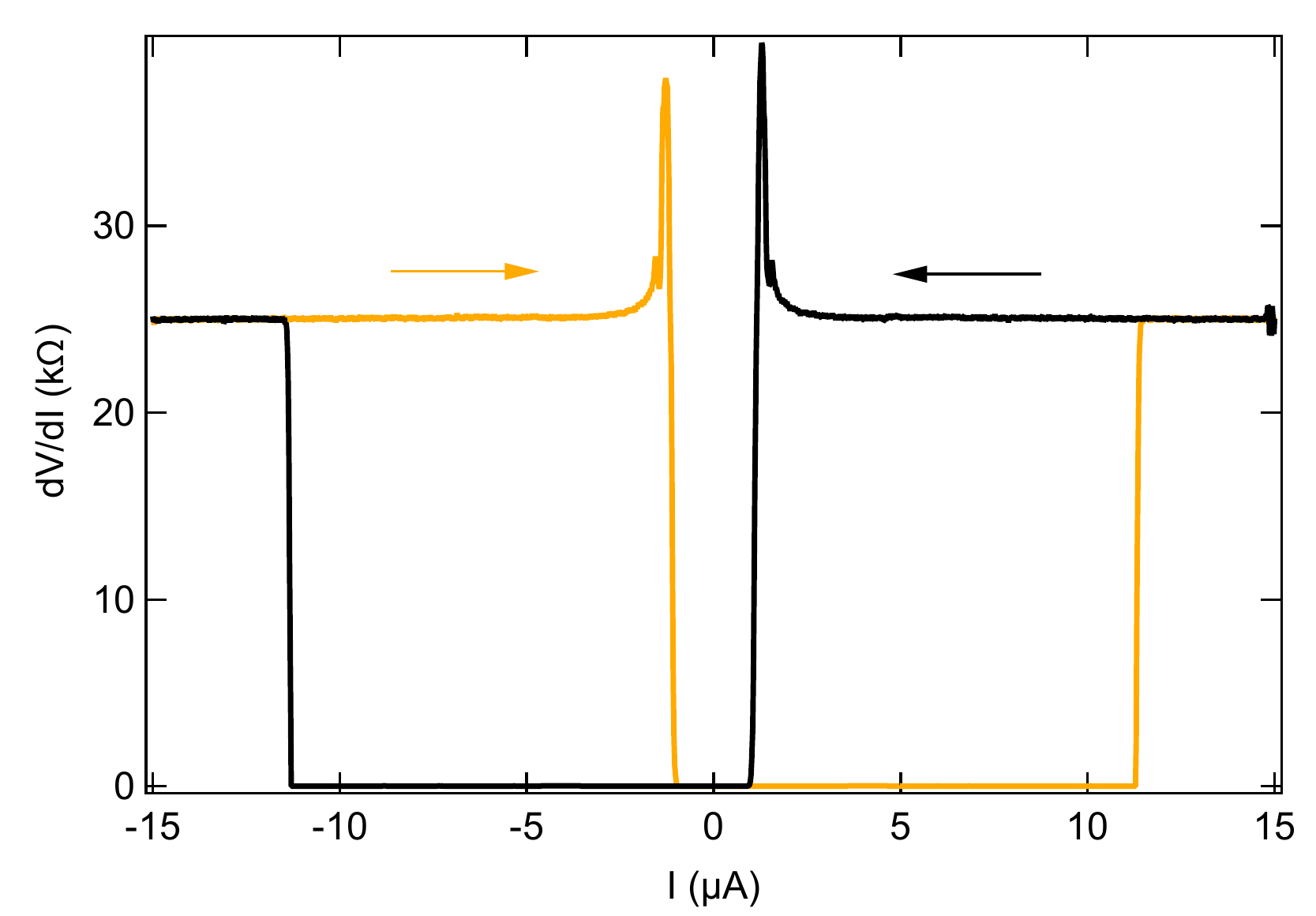}
	\end{center}
	\caption{Differential resistance \textit{vs} current of sample NW2 at very low temperature (T=10mK). The orange curve corresponds to an increase of the current whereas the black one is for decreasing current.}
	\label{FigureIc}
\end{figure}

We finally show in figure \ref{FigureDC}(e) the differential resistance $dV/dI$ of the nanowire NW1 obtained for different applied magnetic fields at $2$~K. At this temperature the curves are non-hysteretic and zero-field critical currents of 4.0/1.3/3.8$~\mu$A were measured for nanowires NW1,NW2 and NW3 respectively.
At lower temperature the supercurrent increases. For NW2 we observed an increase up to 12$\mu$A (critical current density $J_C=62.3$nA.nm$^{-2}$) at $10$~mK  in a highly filtered dilution refrigerator (Fig. \ref{FigureIc}) with a thermal hysteresis behaviour highlighting the increase of electron-phonon relaxation times at low temperature. 

%
%

\section{Non linear resonator}
\label{AppendixB}

\subsection{Temperature dependence}

To access the temperature dependence we use the Mattis-Bardeen (MB) theory \cite{Mattis1958,Barends08} relating resonance frequency and quality factor changes to the temperature and frequency dependent complex conductivity of the material $\sigma=\sigma_1-i\sigma_2$. In this model the resonance frequency reads:
\begin{equation}
f_0= \frac{1}{2 \pi \sqrt{(L_{geo}+L_K)C}} 
\end{equation}
with  $L_K= \mu_0 \lambda \textrm{coth}(d/\lambda)$ being related to the thickness of the wire $d$ and to the London magnetic penetration length $\lambda$. Hence $\lambda$ depends on $\sigma$ as:
\begin{equation}
\lambda=\frac{1}{\sqrt{\mu_0 \omega \sigma_2(\omega,T)}}. 
\end{equation}

The quality factor $Q_{MB}$ is defined as 
\begin{equation}
Q_{MB}=\frac{2}{\alpha \beta(\omega,T)}\frac{\sigma_2(\omega,T)}{\sigma_1(\omega,T)} 
\end{equation}

with $\alpha$ the kinetic inductance fraction and $\beta=1+(2d/\lambda)/sinh(2d/\lambda)$. Finally the real and imaginary part of the conductivity simplifies and can be approximated at low temperature in the local dirty limit with $k_B T,\hbar\omega \ll 2\Delta$ by:

\begin{eqnarray}
\frac{\sigma_1}{\sigma_n}=\frac{4\Delta}{\hbar \omega} e^{-\frac{\Delta}{k_B T}} \textrm{sinh}\left(\frac{\hbar \omega}{2k_B T}\right) K_0 \left(\frac{\hbar \omega}{2k_B T}\right) \\
\frac{\sigma_2}{\sigma_n}=\frac{\pi\Delta}{\hbar \omega} \left[1-2e^{-\frac{\Delta}{k_B T}} e^{-\frac{\hbar \omega}{2k_B T}} I_0 \left(\frac{\hbar \omega}{2k_B T}\right) \right]
\end{eqnarray}

with $I$ and $K$ the modified Bessel functions of the first and second kind respectively and $\sigma_n$ the normal state conductivity of the material.     

\subsection{Power dependence}

Increasing the microwave readout power results in the onset of nonlinear effects due to the nonlinear kinetic inductance of the nanowire. One can write the power-dependent kinetic inductance in terms of the resonator current $I$ as :
\begin{equation}
L_k(I)=L_k(0)[1+(I/I_*)^2+...]
\end{equation}
where $I^*$ sets the scaling of the non-linear effects, in general of the order of the critical current. $L_k(0)$ stands for the low power kinetic inductance. Such non linear term leads to the equations of a Duffing oscillator. To quantitatively tackle this non linear effect in our data one needs to take into account the shift $\delta\omega$ of the resonance frequency due to the non linear kinetic inductance. The shifted resonance then reads 
\begin{equation}
\omega_r=\omega_0+\delta\omega=\omega_0+K n_{ph}
\end{equation}
where we have introduced the Kerr parameter $K$ as the link between the frequency shift and the number of stored photons into the resonator $n_{ph}$. By replacing into the fractional detuning we obtain:
\begin{equation}
x=\frac{\omega - \omega_0 - \delta\omega}{w_0+ \delta \omega}\approx x_0-\delta x= x_0-\frac{K n_{ph}}{\omega _0}
\label{Newdefx}
\end{equation}
with
\begin{equation}
x_0=\frac{\omega - \omega_0}{w_0}
\end{equation}
the normalized detuning in the low power limit. For an energy $E$ stored into the resonator, the shift $\delta x$ is approximately given by:
\begin{equation}
\delta x= \frac{\delta \omega}{\omega_0}=\frac{K n_{ph}}{\omega _0}=-\frac{\delta L}{2 L}=-\frac{\alpha I^2}{2 I_*^2}.
\end{equation}
In order to access the number of photons stored into the resonator $n_{ph}$ as a function of the applied microwave power $P$ at the input of the sample one needs to write the power conservation law relating the lost energy into the resonator $P_{diss}$ with respect to the measurement lines accounted by the scattering parameters corresponding to  wave reflection $S_{11}$ and transmission $S_{21}$. In a hanger resonator the conservation law reads:
\begin{equation}
P_{diss}=P [1-|S_{11}|^2-|S_{21}|^2]
\end{equation}
with $S_{11}=S_{21}-1$. By replacing in this equation the $S_{21}$ formula, we find:
\begin{equation}
P_{diss}=P \left[\frac{2 Q_t^2}{Q_i Q_c} \frac{1+4\frac{Q_c^2 Q_t}{Q_c-Q_t} u x}{1+4 Q_t^2 x^2}\right].
\end{equation}
Inserting in this equation the definition of the internal quality factor $Q_i=E\omega_r/P_{diss}= n_{ph} \hbar\omega_r^2/P_{diss}$ so that $Q_i\approx n_{ph} \hbar\omega_0^2/P_{diss}$ gives access to the number of photons in the resonator as a function of the applied power:
\begin{equation}
n_{ph}=\frac{2 Q_t^2 P}{Q_c \hbar \omega_0^2} \frac{1+4\frac{Q_c^2 Q_t}{Q_c-Q_t} u x}{1+4 Q_t^2 x^2}.
\end{equation}
Incorporating this expression into equation \ref{Newdefx} leads to:
\begin{equation}
x= x_0-\frac{K n_{ph}}{\omega _0} = x_0 - K \frac{2 Q_t^2 P}{Q_c \hbar \omega_0^3} \frac{1+4\frac{Q_c^2 Q_t}{Q_c-Q_t} u x}{1+4 Q_t^2 x^2}.
\label{newdefxbis} 
\end{equation}
Introducing the variables $y=Q_t x$, $y_0=Q_t x_0$ and $v=Q_t u$, the non-linear parameter
\begin{equation}
a_{NL}= -\frac{2 K Q_t^3}{Q_c \hbar \omega_0^3} P
\label{aNL}
\end{equation}
and simplifying the expression by noting 
\begin{equation}
c=\frac{Q_c^2 Q_t}{Q_c-Q_t}
\end{equation}
we can rewrite in a more convenient form the expression \ref{newdefxbis} as:
\begin{equation}
y=y_0 + \frac{a[1+4cvy]}{1+4y^2}.
\label{yequation}
\end{equation}
It consists in a third degree polynomial which admits in general three solutions that are post-selected to match the measurement sweep direction. By solving this equation \ref{yequation} it is possible to calculate the power-dependent frequency detuning $x$ and fit the experimental data using the $S_{21}$ formula in the main text. Doing so allows one to plot as a function of the applied microwave power $P$, the non linear term $a_{NL}$, $Q_i$, $Q_c$ and $Q_t$. As expressed in formula \ref{aNL}, $a_{NL}$ should be a linear function of the power with a slope $s$ related to the non-linear Kerr parameter $K$ such that 
\begin{equation}
K=-\frac{Q_c \hbar \omega_0^3}{2 Q_t^3} s.
\label{Kformula}
\end{equation}
\begin{figure}[htbp]
	\begin{center}
		\includegraphics[width=7.5cm]{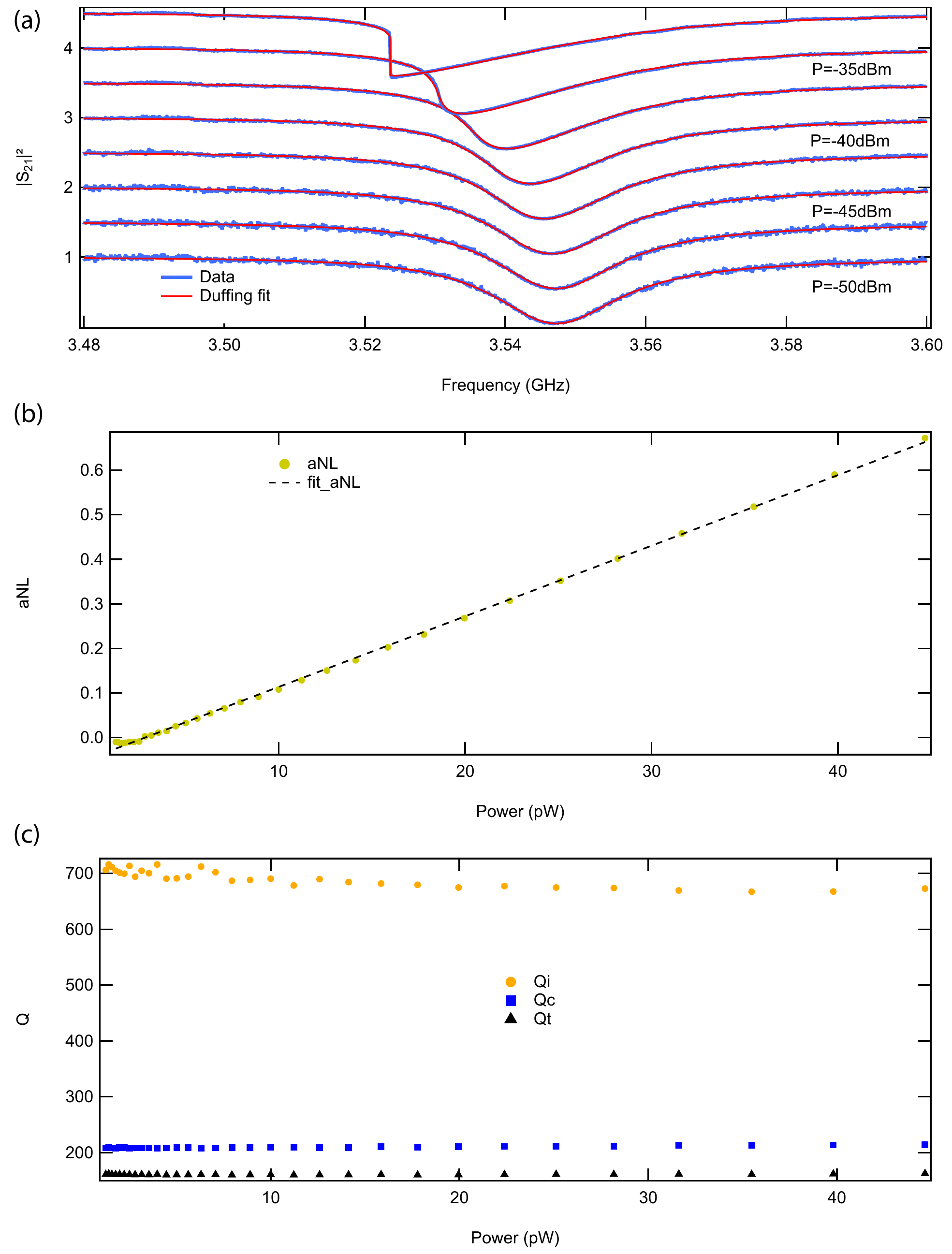}
	\end{center}
	\vspace{-20pt}
	\caption{(a) Power dependence of the normalized transmission spectra near the microwave resonance of the lumped resonator at 1.5K. (b) Power dependence of the aNL term allowing us to extract (see text) a Kerr parameter $K_{W,He}=74$~Hz/photon at $3.5$GHz. (c) Extracted quality factors as a function of the microwave power. The Q's seem to be rather stable with respect to Power with a maximum internal quality factor of 700.}
	\label{Figure55}
\end{figure}

\subsection{Magnetic field dependence}
The dependence versus magnetic field was done on a lumped element resonator with a $30\mu$m long, $80$~nm wide, $25$~nm thick W nanowire with a resonance frequency around $3.55$GHz at $T=1.55$K in a pumped He bath. We extracted from the power dependence of the resonance shown in figure \ref{Figure55} a non-linear Kerr parameter $K_{W,He}=74$~Hz/photon with an internal $Q_i$ factor of $700$. 

\begin{figure}[htbp]
	\begin{center}
		\includegraphics[width=8.5cm]{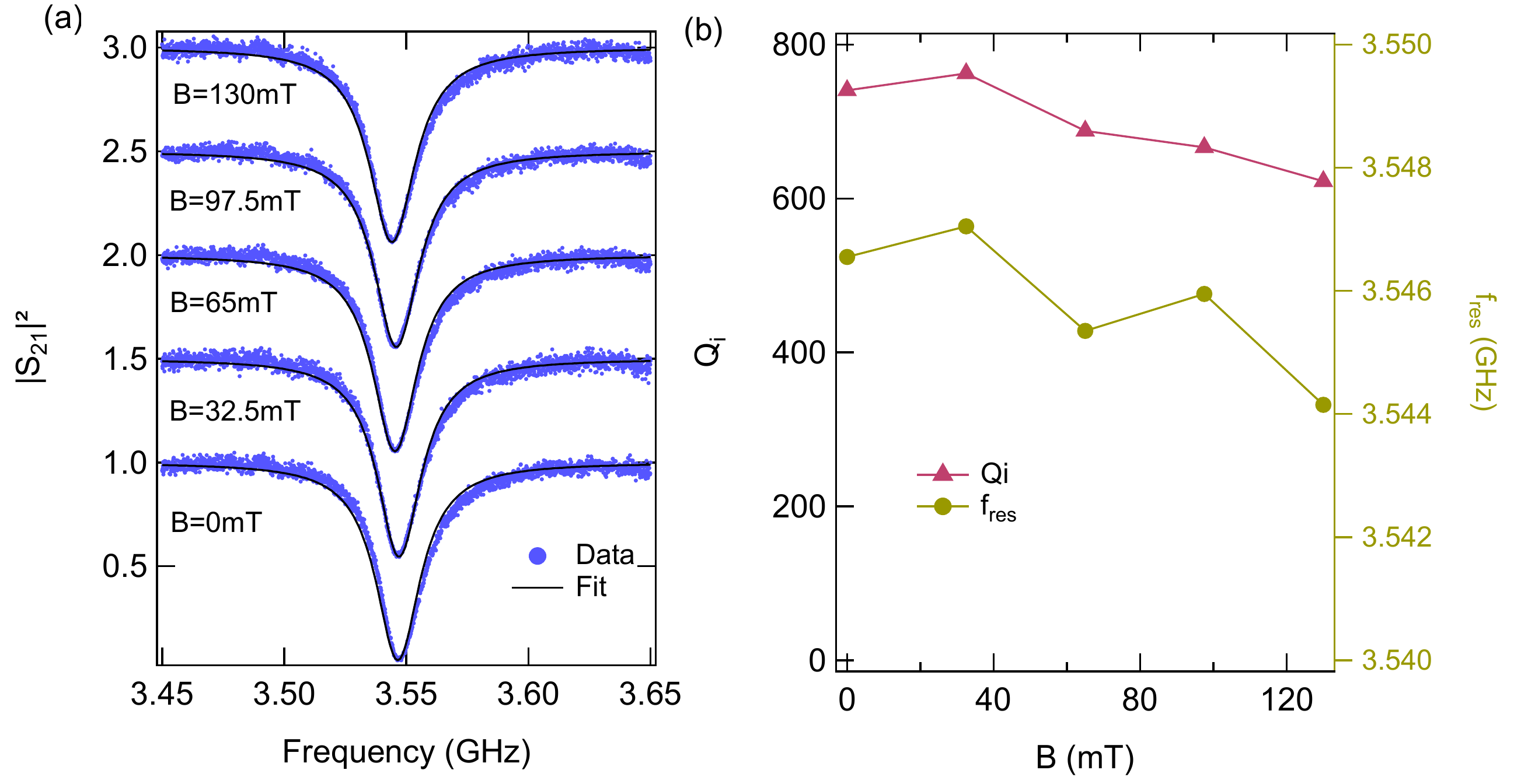}
	\end{center}
	\vspace{-20pt}
	\caption{(a) In plane magnetic field dependence of the normalized transmission spectra of the lumped resonator at $1.55$K with fits using the $S_{21}$ formula. (b) Evolution of $Q_i$ and $f_{res}$ \textit{vs} magnetic field.}
	\label{Figure6}
\end{figure}

As such the hybrid lumped resonator presented here is an interesting candidate to realize relatively high sensitivity photon detection experiments at a very moderate temperature such as the one of a pumped He bath.
Because of the high critical fields of both materials used for this experiment, Nb and W, the resonator should be immune to applied magnetic field. We thus measured the microwave resonance up to $130$~mT. As expected we observe (see figure \ref{Figure6}) small changes for $Q_i$ ($<10\%$) and $f_{res}$ ($<0.05\%$) which makes this hybrid resonator interesting to study mesoscopic devices where the spin degree of freedom needs to be addressed.

\begin{figure}[htbp]
	\begin{center}
		\includegraphics[width=7.5cm]{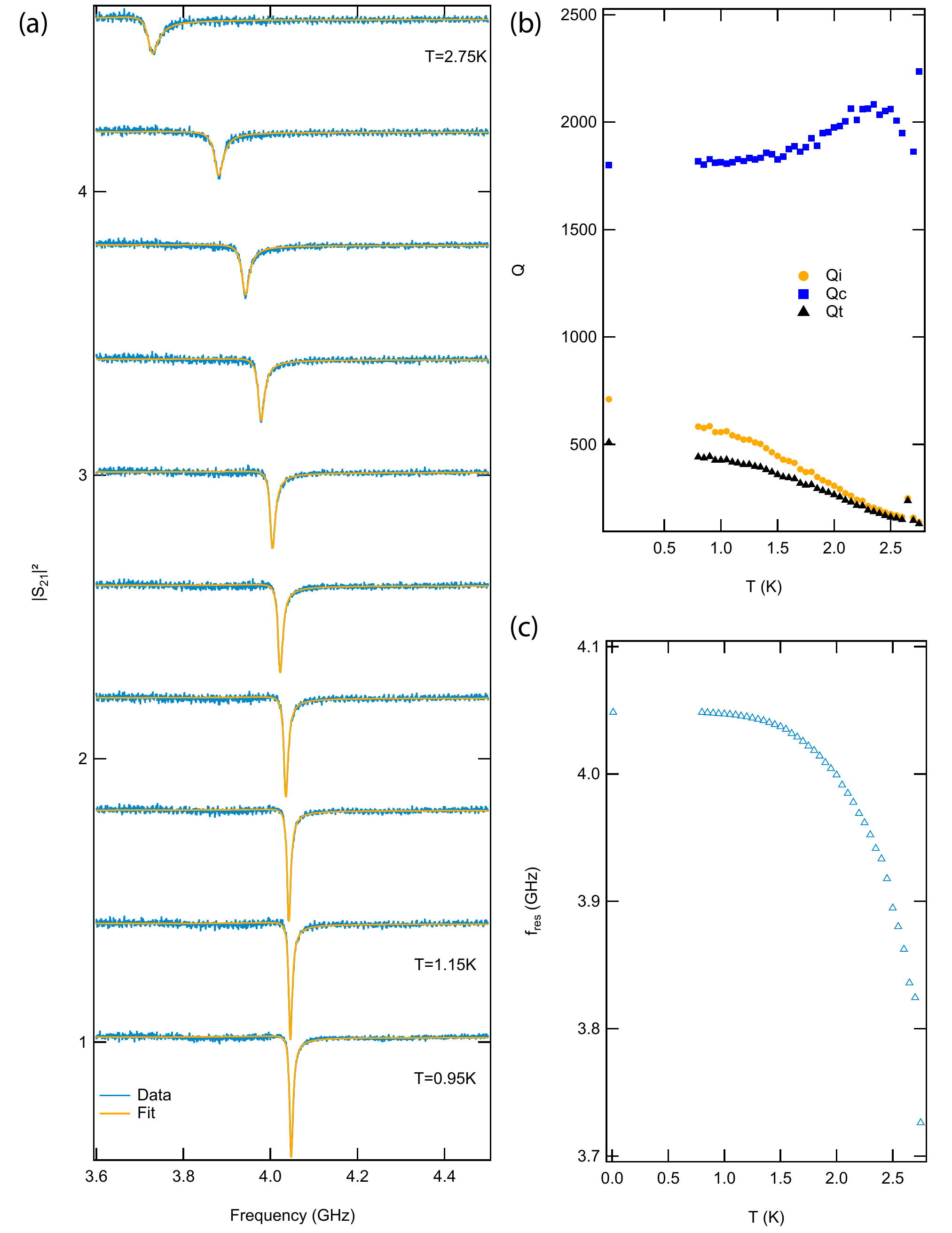}
	\end{center}
	\vspace{-20pt}
	\caption{(a) Temperature dependence of the normalized transmission spectra of the CPW resonator. (b) Extracted quality factors \textit{vs} temperature. (c) Temperature dependence of the resonance frequency.}
	\label{Figurelambdasur4}
\end{figure}
\subsection{CPW $\lambda/4$ resonator}

As mentioned in the main text we also studied a CPW type of resonator. The temperature dependence of the resonance can be found in figure \ref{Figurelambdasur4}. The internal quality factor at $10$~mK was $Q_i=710$ and the Kerr parameter was $K_{W,He}/2\pi=68$~Hz/photon.

\end{appendix}


\begin{thebibliography}{99}
\bibitem{footnote} Each resistance measurement was done in a two probe configuration and a contact resistance was substracted from the raw data.
\end{thebibliography}
\end{document}